\documentclass[12pt,preprint]{emulateapj}

\def\msol{\mbox{M}_\odot}
\def\zsol{\mbox{Z}_\odot}

\def\minf{m_{\rm inf}}
\def\msup{m_{\rm sup}}
\def\mvol{\mbox{M}_\odot\, \mbox{pc}^{-3}}
\def\mwd{\mbox{m}_{WD}}

\def\mh{\,\mbox{[M/H]}}

\def\mvol{\mbox{M}_\odot\, \mbox{pc}^{-3}}

\def\mwd{\langle m_{\rm WD}\rangle}
\def\beq{\begin{equation}}
\def\eeq{\end{equation}}

\def\simgr{\,\hbox{\hbox{$ > $}\kern -0.8em \lower 1.0ex\hbox{$\sim$}}\,}
\def\simle{\,\hbox{\hbox{$ < $}\kern -0.8em \lower 1.0ex\hbox{$\sim$}}\,}

\lefthead{Chabrier}
\righthead{The glow of primordial remnants}

\begin{document}

\title{The glow of primordial remnants}
\author{G. Chabrier}
\affil{Ecole Normale Sup\'erieure de Lyon, C.R.A.L. (UMR CNRS 5574), 69364 Lyon, France\\
and Erskine visitor, Dpt. of Physics \& Astronomy, University of Canterbury, Christchurch, New Zealand}

\begin{abstract}

We determine the expected surface brightness and photometric signature of a white dwarf remnant population, issued from primordial low-mass stars formed at high redshifts, in today galactic halos. We examine the radial dependence of such a contribution as well as its redshift dependence. Such a halo diffuse radiation is below the detection limit of present large field ground-based surveys, 
but should be observable with the HST and with the future JWST project. Since the surface brightness does not depend on the distance,
the integration of several galactic dark halos along the line of sight  will raise appreciably the chances of detection. Both the detection or the non-detection of such a remnant diffuse radiation within relevant detection limits offer valuable information on the minimum mass for star formation in the early universe and on the evolution of the stellar initial mass function.
\end{abstract}


\section{Introduction}

Modern theories of structure formation in the early universe suggest that galaxies have formed out of
a primordial generation of stars, with zero-like metal abundances, the so-called Population III
stars. Our understanding of the general properties of Pop III stars has improved significantly over
the past recent years (see Barkana \& Loeb 2001, Bromm \& Larson 2004 for recent reviews). The mass distribution of
this first generation(s) of stars, however, remains ill-determined. According to recent hydrodynamical
models (Abel, Bryan \& Norman 2000, Bromm, Copi \& Larson 1999, 2002) the first stars are very massive ($\simgr 100\,\msol$),
primarily because of the poor efficiency of the dominant cooling mechanism, due to molecular
hydrogen (H$_2$) rotational transitions. Other calculations, however, find that the formation of Pop III stars following the very first-generation stars may extend
to the $\sim 1$-10 $\msol$ range, depending on the protogalactic gas cloud conditions (Nakamura \& Umemura 2001, 2002),  the effect of the intense radiation field (Omukai \& Yoshii 2003) or on the efficiency of metal-enriched (Pop II.5) star formation due to supernovae (Mackey, Bromm \& Hernquist 2003, Salvaterra, Ferrara \& Schneider 2003). In that case, primordial star formation may have led to an early generation of extremely or totally (in the absence of mixing of  the ejected material in the SN-driven shell in the ISM) metal-depleted, long-live low-mass  stars. The minimum
mass for star formation in the early universe thus remains uncertain. This
minimum mass, however, is likely to decrease with time, as (i) the virial temperature of massive collapsing halos will exceed $T_{vir}>10^4$, inducing radiative line-cooling from atomic hydrogen, (ii)
the minimum ambient temperature, set up by the cosmic background radiation, will decrease and so
will the thermal Jeans mass, and (iii) a small fraction of metals will be processed
in the primordial short-lived stars, yielding a much more efficient cooling mechanism (Clarke \& Bromm 2003, Bromm \& Loeb 2003). When the transition between Pop III+II.5, hereafter defined generically
as our "primordial star generation", to well identified Pop II-like stars is occurring remains unknown,
depending on ill-determined factors such as the effect of the ambient UV radiation on the
main coolants or the efficiency
of the medium heavy element enrichment due to the first SNII generations.

The evolution of the initial mass function (IMF) from the (nearly-zero $Z$) Pop III to (low-$Z$) Pop II may in fact correspond to transition from a high-mass to a low-mass fragmentation
mode of star formation, illustrating the evolution of the IMF with structure formation in the universe.
The recent discovery of a very metal-depleted ($[Fe/H]=-5.3\pm 0.2]$) red giant ($m\approx 0.8\,\msol$) in our Galactic halo (Christlieb et al. 2002) indeed points to such a low-mass mode of
primordial star generation. Moreover, the metal enrichment of clusters of galaxies suggests an IMF 
at high redshift with
an average stellar mass larger than for present-day conditions, in the range $\sim 1$-$10\,\msol$
(Portinari et al. 2004 and references therein).
Such an evolution bears important consequences
on the early history and the subsequent evolution of the universe. While the UV radiation and $\alpha$-element
nucleosynthesis will be dominated by
the more massive objects, a primordial, high-redshift stellar population extending to
masses below about 10 $\msol$ will (i) extend the nucleosynthetic products to C, N and Fe enriched elements as well as s-processed elements and (ii) produce a
population of white dwarfs (WDs) sequestering part of the baryons. 
The quest for the direct detection of these remnants remains for now inconclusive, for present surveys do not reach deep enough magnitude or
have a too small field of view to reach robust conclusions. An alternative observable signature of the presence of a primordial remnant population in today galactic halos would be their contribution to these halo surface brightness. N-body simulations (White \& Springel 2000) suggest that the primordial star
clusters might survive disruption during halo merging preceding galaxy formation and, through
loss of kinetic energy by dynamical friction, end up in the central bulge, with low velocity.
If, however, the low-mass stars are scattered by violent relaxation or tidal disruption of the cluster during halo merging, the remnants of these early stars would be now incorporated into galactic external halos.
Which of these two processes dominates remains to be determined unambiguously and depends on various factors such as
the mass ratio and the compactness of the mergers or the efficiency of the SN-driven winds in the clusters. In the present paper we will assume that the primordial remnant population follows the isothermal density distribution
of the dark halo. If these remnants are located in the central bulge or spheroid, their surface brightness
will be largely dominated by the main sequence (M-dwarfs) contribution. In that case only spectroscopic
and high-proper
motion detections of individual objects will allow their identification as PopIII remnants.
The nature of these primordial remnants (black holes, neutron stars or white dwarfs) carries
the signature of the minimum mass of the IMF at the early epoch of star formation in the universe.
Since the surface brightness does not depend on the distance (see below), the integration
along the line of sight of several galactic dark halos adds up each contribution, yielding eventually a
detectable signature. 
Both the observation or the {\it non} observation of such a stellar relic radiation would bring important information  on the low-mass end of the stellar mass spectrum and thus on the dominant process of star formation in the early universe.
This is the aim of the present paper to examine this issue.

\section{Variation of the initial mass function}

\begin{figure}
\plotone{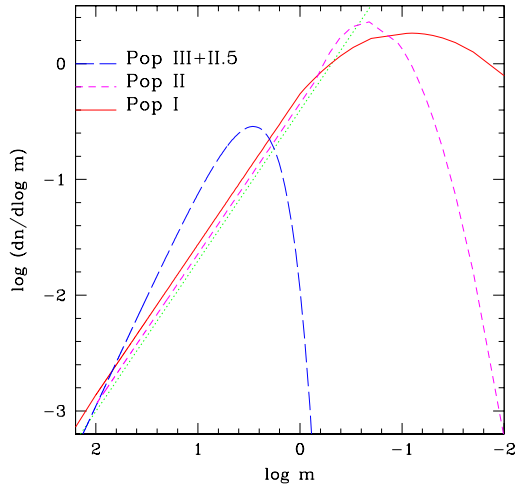}
\caption{Illustration of the possible variation of the IMF with redshift and characteristic stellar populations.
The PopI, PopII and PoPIII+II.5 IMFs correspond to Table 1, Table 2 and eqn.(24) of Chabrier (2003), respectively. The dotted line corresponds to the Salpeter IMF. All IMFs normalized to $\int_{0.01}^{100}m\,(dN/dm)\,dm=1$.
\label{f1.ps}}
\end{figure}

As mentioned above, several arguments suggest that the IMF was more top-heavy at high redshift than
in present day environment. Figure \ref{f1.ps} illustrates the possible variation of the IMF from PopIII+II.5,
formed at high-redshift  ($z\simgr 10$ or so)
with very low metal enrichment ($\mh \ll -2$), to PopII stars,
with $\mh\sim -2.0$ to -1.0, characteristic of the spheroid and globular cluster population,
to solar metallicity PopI stars representative of the disk population. These IMFs correspond to eqns. (17), (20) and (24) (see also Tables 1 and 2)
of Chabrier (2003), respectively. These IMFs are best described by a power-law at large masses
rolling down to a lognormal form below some characteristic mass, a behaviour expected from turbulence-driven fragmentation (Padoan \& Nordlund 2002),
but can be approximated by the following form, as initially suggested by Larson (1986) :

\begin{eqnarray}
\phi(m)\propto m^{-\alpha}\,exp\Bigl[-({m_0\over m})^{\beta}\Bigr]
\label{imf}
\end{eqnarray}

As mentioned above, it is unclear whether a WD-dominated primordial IMF such as the one displayed in Figure \ref{f1.ps} is representative of the primordial PopIII+II.5 stellar generation. However, as argued above,
different calculations suggest that the next to very first stellar generation may extend into this domain.
Moreover, the aforementioned recent observation of a 0.8 $\msol$ red giant with a strong iron depletion ($[Fe/H]=-5.3$) suggests indeed that the primordial IMF quickly evolved into the WD progenitor mass range. 
In order (i) to maximize the
contribution from the putative WD-progenitor population and thus estimate the maximum constraint on this population, and (ii) to examine the dependence of the diffuse radiation upon the IMF and thus
WD mean mass $\mwd$,
we chose for illustration the IMF suggested by Chabrier, Segretain \& M\'era (1996), with
$m_0=2.0\,\msol,\,\alpha=5.1,\,\beta=2.2$ (their IMF1), yielding $\mwd=0.62\,\msol$ and $m_0=2.7\,\msol,\,\alpha=5.7,\,\beta=2.2$ (their IMF2), which yields $\mwd=0.68\,\msol$. 

\section{Extragalactic background light}

An important constraint on the presence of such a primordial remnant population is the contribution of its progenitors to the energy content of the universe, i.e.
to the cosmic background light. This contribution was first estimated
by Madau and Pozzetti (2000).  These authors, however, used solar-metallicity or only slightly
metal-depleted (1\% $\zsol$) stellar evolution models to evaluate this contribution. For a given mass, zero-metallicity stars are brighter and thus put more severe constraints on the related energy background. More recently Santos, Bromm
\& Kamionkowski (2002) carried out
more detailed calculations. They considered, however, only very massive stars ($m \simgr 300 \,\msol$)
and found out that the spectral properties of these stars are essentially independent of mass,
when normalized to unit stellar mass, so that the integrated spectrum of such a population depends
only on its total mass, not on the IMF. This result does not hold anymore for lower mass stars, and
has been shown recently to be partially incorrect even for the highest mass stars (Schaerrer 2002).
More recently, Salvaterra \& Ferrara (2003) and Magliocchetti, Salvaterra \& Ferrara (2003) have shown that Pop III stars formed at $z\simgr 9$ can contribute appreciably, possibly entirely to the detected residual (i.e. after galaxy contribution subtraction) NIRB radiation, without major conflict with the IGM observed metallicity limits.
All these calculations suggest that a substantial fraction of the collapsed (virialized) baryons, $\sim 10$ to 60\% depending on the primordial IMF and the redshift range of formation, may have been
turned into this primordial generation of stars. 

This point is illustrated below.
Following Madau \& Pozzetti (2000), we calculate the bolometric extragalactic background light per steradian $I_{EBL}$ observed
at Earth today ($z=0$), due to a burst of formation of primordial stars of zero-metallicity at time $t_F$, i.e. redshift $z_F$.  Since zero-metallicity stars are brighter and hotter than their metal-enriched counterparts,
they evolve more quickly and the redshift dependence of the produced light is likely to differ from the estimate of Madau \& Pozzetti (2000). A star-burst approximation is reasonable since primordial star formation is believed to have occurred somewhere between $z_F\sim 20$-30, when small scale halos were formed from primordial fluctuations, and $z_F\sim 6$, about the end of the reionization. In the
standard cosmological model adopted in the present calculations, $(h,\Omega_m,\Omega_\Lambda)=(0.70,0.30,0.70)$, this spans $\sim\,10^8$ yrs, a fairly short time compared with the age of the universe today, $t_h=13.7$ Gyr. 
The $I_{EBL}$ is given by:

\begin{eqnarray}
I_{EBL}={c\over 4 \pi} \int_{t_F}^{t_h} {\rho_{bol}(t) \over 1+z(t)} dt
\end{eqnarray}

\noindent where 

\begin{eqnarray}
\rho_{bol}(t)=\int_0^t L(\tau) \dot \rho (t-\tau) d\tau
\label{rhobol}
\end{eqnarray}

\noindent is the total bolometric luminosity density
 for a stellar formation rate per comoving volume $\dot \rho$ and

\begin{eqnarray}
L(\tau)=\int_{m_{inf}}^{m_{sup}} l(m,\tau) \phi(m) dm
\end{eqnarray}

\noindent is the luminosity produced by a stellar population of same age $\tau$, from
a minimum mass $m_{inf}$ to a maximum mass $m_{sup}$. 
We use the
zero-metallicity evolutionary sequences of Marigo et al. (2001), with $m_{inf}=0.1\,\msol$ and $m_{sup}=1000\,\msol$.
The term $l(m,t)$ denotes the luminosity of a star with mass $m$ at age $\tau$, and $\phi(m)=dN/dm$
is the IMF normalized such that the total stellar mass $M=\int_{m_{inf}}^{m_{sup}} m \phi(m) dm=1$.
For a burst SFR, eqn.(\ref{rhobol}) reduces to $\rho_{bol}(t)=\rho_{m_\star}\, L(t-t_F)$, where $\rho_{m_\star}$ is the mass density under the form of stars.

\begin{figure}
\plotone{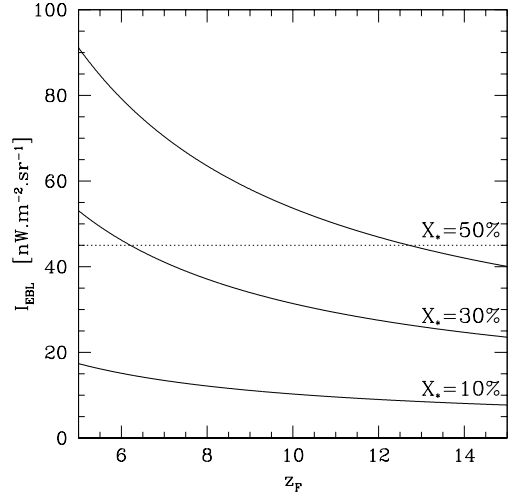}
\caption{Extragalactic background light received at Earth today due to a burst of zero-metallicity stars
formed at redshift $z_F$ with IMF1 and containing a mass fraction $X_\star$ of the nucleosynthetic baryon density. The dotted line indicates the lower limit of the observed light.
\label{f2.ps}}
\end{figure}

\noindent 
Figure \ref{f2.ps} displays $I_{EBL}(z_F)$ expected for the aforedescribed population of primordial stars as a function of redshift of formation $z_F$ for various mass fractions $X_\star(z)$ of baryons under the form of stars at redshift $z$ :

\begin{eqnarray}
\Omega_\star(z)={\rho_{m_\star}(z)\over \rho_c}=X_\star(z)\times \Omega_B,
\end{eqnarray}

\noindent where $\rho_C=3H_0^2/8\pi G$ is the universe critical density and $\Omega_B$ is the baryon density in units of $\rho_C$. The
observed (frequency-integrated) brightness of the cosmic background radiation today is
$I_{EBL}\simeq 100$ nW m$^{-2}$ sr$^{-1}$, with a lower limit $I_{EBL}\simeq 45$ nW m$^{-2}$ sr$^{-1}$ (Hauser \& Dwek 2001), indicated by the dotted line in the figure.

These calculations are by no means as accurate as the detailed calculations mentioned earlier.
In particular they do not incorporate other contributions like e.g. normal galaxies. However, they illustrate the fact that
a significant mass fraction of primordial stars, possibly as much as $\Omega_\star=0.5\times \Omega_B$ can be formed at redshift $z\simgr 5$ and contribute a diffuse background
light today well below the observed lower limit.
Since the IMF extends in the WD-progenitor range and since WD progenitors return about 70\%
of their mass to the ISM with the IMF (\ref{imf}) (assuming a similar WD-mass vs progenitor-mass relationship as for the disk WDs), this implies that, in principle, as much as $\sim 15\%$ of the nucleosynthetic baryon density can be trapped in a population of remnants from
a primordial generation of low-mass stars. 

This is of course a large overestimate of the true fraction of baryons possibly processed into PopIII+II.5
remnants. This fraction can be estimated as follows. The fraction of collapsed baryons at $z\sim 10$ is
$F_b\approx 0.1$ (Haiman \& Loeb 1997\footnote{Haiman \& Loeb (1997) simulations are based on
Press-Schechter SCDM and different cosmological parameters but modern simulations do not seem to
change drastically these results.}, Abel et al. 1998, Bromm et al. 1999). The star formation efficiency in today star-forming regions range from $f_\star \sim 0.1$ to 0.3 and, as mentioned above, may reach $f_\star \sim 0.6$ at high redshift (Salvaterra \& Ferrara (2003), Magliocchetti et al. 2003). Given an estimate of $f_{esc}\sim 0.1$
from the observation of
local galaxies,
where $f_{esc}$ is the escape fraction of ionizing photons within these protogalaxies, this is
consistent with the constraint $f_\star\times f_{esc}\simle 0.1$ from the WMAP Thomson optical depth and the lack of H I Gunn-Peterson through at $z\simle 6$ (Fig. 5 of Wyithe \& Loeb 2003). This yields a fraction of BBN bayons $X_\star =f_\star \times F_b(z)\approx 1$-10\% possibly processed into PopIII+II.5 stars at large redshift. Recent studies suggest that the direct detection of the radiation of these primordial star clusters, responsible for the universe reionization, might be in reach of the JWST or even the HST missions (Venkatesan \& Truran 2003, Stiavelli, Fall \& Panagia 2004). On the other hand,
with the type of IMF mentioned earlier,
about 30\% of these baryons might be under the form of WD remnants
in galactic halos today.
The photometric signature of such a remnant population in galactic halos is examined below.

\section{Diffuse emission of Population III white dwarfs}

The mass fraction $X_{WD}=\rho_{WD}/\rho_{h_\odot}$, where $\rho_{h_\odot}\simeq 9\times10^{-3}\,\msol$ pc$^{-3}$ is the local dark matter density, of WD remnants in the Milky Way halo,
issued from the primordial low-mass star generation considered in the previous section, remains presently undetermined.  The
value inferred from microlensing experiments has been steadily decreasing over the years,
and is now estimated to be between $X_{WD}=0$ and $\sim$20\% (Alcock et al. 2000, Afonso et al. 2003), with $X_{WD}< 5\%$ at the 95\% C.L. from the EROS team. A $X_{WD}\sim 2\%$ population of these halo WDs
has been claimed to have been identified recently from kinematic and spectroscopic observations (Oppenheimer et al. 2001). Although this result remains highly controversial (see e.g. Reid, Sahu \& Hawley 2001, Reyl\'e, Robin \& Cr\'ez\'e 2001, Flynn, Holopainen \& Holmberg 2003),
a detailed analysis of the age and kinematic properties of these objects seems to confirm that part of the sample does belong to the Galactic halo (Hansen 2001). Present surveys, however, either have too small field of views or do not reach faint enough magnitudes to detect the peak of the halo WD
luminosity function and thus the majority of such a halo WD population remains presently out of reach (Chabrier 1999a Figure 1). 
As pointed out by Chabrier (1999a), the only robust constraint yielding  an upper limit for $X_{WD}$ 
(besides the maximum baryonic fraction estimated in \S 3), arises from stellar nucleosynthesis of helium and heavy element production. Recent detailed calculations seem to exclude a value
$X_{WD} \simgr$1\% to 10\% (Fields, Freese, \& Graff 2000, Brook, Kawata \& Gibson 2001).
In any events, a primordial, high-velocity ($v_\perp \simgr 100$ km s$^{-1}$) WD population with $X_{WD}\simgr 1\%$, as suggested by Oppenheimer et al. (2001), would exceed significantly the WD population expected in the Galactic spheroid, $\rho_{WD}\simeq (1.8\pm 0.8)\times 10^{-5}\,\mvol$, i.e. $X_{{WD}_{sph}}\approx 0.2\pm 0.1\%$ (Gould, Flynn \& Bahcall 1998, Chabrier 2003 Table 3), and would thus imply a primordial IMF different from the one characteristic of today environments.
At any rate, the identification of any halo WD population would bring precious information about the minimum mass for star formation at early stages of galactic formation. It is thus interesting to explore the expected diffuse emission of such a population.

Following the formulation of Adams \& Walker (1990), we calculate the radiation signature of a WD population with mass fraction $X_{WD}$ in an isothermal halo. 
The specific luminosity per unit mass $\langle \Gamma_\nu \rangle_\lambda$ in a broadband filter characterized by central wavelength $\lambda$ and width $\Delta \lambda$ of a WD population
obeying the distribution given by an IMF $dn/dm_{WD}$, reads:

\begin{eqnarray}
\langle \Gamma_\nu \rangle_\lambda={ \langle \L_\nu \rangle_\lambda \over \langle m_{WD} \rangle }
\end{eqnarray}

\noindent with

\begin{eqnarray}
\langle L_\nu \rangle_\lambda = { \int_{\minf}^{\msup} \bar L_{\nu_\lambda} (m_{WD},\tau)  {dn\over dm_{WD}} \, dm_{WD}
 \over \int_{\minf}^{\msup} {dn \over dm_{WD}} \, dm_{WD}  }
\end{eqnarray}

\noindent and
\begin{eqnarray}
\langle m_{WD} \rangle = { \int_{\minf}^{\msup} {dn \over dm_{WD}} m_{WD} \, dm_{WD} \over  \int_{\minf}^{\msup} {dn \over dm_{WD}} \, dm_{WD}}
\end{eqnarray}

\indent Here, $(\minf,\msup)=(0.5\,\msol, 1.2\,\msol)$ denote respectively the minimum and maximum masses for a
standard WD mass spectrum, whereas ${\bar L_{\nu_\lambda}(m_{WD},\tau)}=4\pi R_{WD}^2 (m_{WD},\tau) \langle {\mathcal F_{\nu}}\rangle_\lambda (m_{WD},\tau)$ is the luminosity per unit frequency averaged over the filter $\lambda$ for a WD of mass $m_{WD}$ and age $\tau$, emitting at its
surface a flux $\mathcal F_{\nu}$.

The surface brightness $I_\nu$ and the flux density at frequency $\nu$ per unit mass $f_{\nu}=\Gamma_\nu/4\pi s^2$ observed at  Earth from such a population emitting at distance $s$ in a galactic halo of density profile $\rho [{\bf r}(s)]$, where $r$ and $s$ denote distances from the Galactic centre and from the Sun, respectively, 
reads:

\begin{eqnarray}
I_\nu={ f_\nu \over \Omega } &=& {X_{WD} \over \Omega }\int_{V}{\Gamma_\nu \over 4\pi s^2}\, \rho [{\bf s}(b,l)] dV \nonumber \\
&=& X_{WD}\; {\Gamma_\nu \over 4\pi}  \int_0^{s_{max}} \rho [ s(b,l)] d s
\label{bright}
\end{eqnarray}

The apparent and absolute magnitudes in the filter $\lambda$ read, respectively :

\begin{eqnarray}
m_\lambda &=&-2.5\,\log \langle f_{\nu}\rangle_\lambda + C_\lambda \\
M_\lambda &=&-2.5 \,\log \bar L_{\nu_\lambda} + 2.5\,\log(4 \pi [10 {\rm pc}]^2) + C_\lambda 
\end{eqnarray}

\noindent where $\langle f_{\nu}\rangle_\lambda=\int_0^\infty f_\nu S_\lambda(\nu) \, {\rm d} \nu / \int_0^\infty S_\lambda(\nu) \, {\rm d} \nu$ is the flux distribution in the broadband filter characterized by the transmission function $S_\lambda (\nu)$, and $C_\lambda$ denotes the normalization constant plus zero-point correction in the filter. The present calculations are done in the standard filters, with
magnitudes normalized to Vega (Bessell \& Brett 1988, Allen 1991).
The absolute magnitudes $M_\lambda (m_{WD},\tau)$ and radii $R_{WD}(m_{WD},\tau)$ of the WDs are taken from the cooling sequences
of Chabrier et al. (2000)\footnote{Available at http://perso.ens-lyon.fr/gilles.chabrier/WD2000.tar}, based on the spectral energy
distributions of Saumon \& Jacobson (1999). This describes the evolution of so-called DA WDs, with hydrogen rich atmospheres, the most likely population, given WD accretion rates (Chabrier 1999b). At any rate, DA WDs, for a given age, will be order of magnitude brighter than helium-rich atmosphere so-called DB WDs, and will thus dominate the diffuse radiation flux. Note that the age $\tau$ of the WD includes both the WD cooling time $t_{cool}$ and the main sequence lifetime $t_{MS}$ of the progenitor of mass $m$, with the condition that the WD population today, at halo age $t_h$, must obey the condition (Chabrier et al. 1996, Chabrier 1999a):

\begin{eqnarray}
t_h=t_{MS}(m)+t_{cool}(L,m_{WD})
\end{eqnarray}

\subsection {Milky Way halo}

The surface brightness (\ref{bright}) is rewritten:

\begin{eqnarray}
I_\nu=X_{WD}\, {\Gamma_\nu  \over 4\pi} \rho_{h_\odot} R_\odot {\mathcal I} (b,l)
\label{imw}
\end{eqnarray}

\noindent where
$R_\odot=8.5$ kpc is the Galactocentric position of the Sun, and 
${\mathcal I} (b,l)=(\rho_{h_\odot} R_\odot)^{-1}\int_0^{s_{max}} \rho [s(b,l)]ds$ denotes the angular integration along the line of sight. 

The density distribution of a flattened isothermal halo with galactocentric cylindrical coordinates $(R,z)$ reads :

\begin{eqnarray}
\rho(R,z)&=& \rho_{h_\odot}\, ({ \sqrt {1-q^2} \over q\, {\rm Arccos}\, q})\,{ R_\odot^2 + a^2 \over R^2 + a^2 + z^2/q^2 }\nonumber \\
&=&{v^2_{rot}\over 4\,\pi\,G}\, ({ \sqrt {1-q^2} \over q\, {\rm Arccos}\, q})\,{ 1 \over R^2 + a^2 + z^2/q^2 }
\label{rho}
\end{eqnarray}

\noindent where $a\simeq 5$ kpc denotes the halo core radius and $q\simeq 0.7$ the oblateness.

For such a density distribution, the integral (\ref{imw}) can be evaluated analytically: 

\begin{eqnarray}
{\mathcal I} (b,l) &=&
{\alpha^2 +1 \over \xi} \Bigl\{ \tan^{-1} ({\mu \over \xi}) \nonumber \\
&-&  \tan^{-1}( { \mu - Y_{max} (1+Q^2\sin^2 b) \over \xi }) \Bigr\} 
\label{angle}
\end{eqnarray}

\noindent with $\mu=\cos b \, \cos l$, $\alpha = a/R_\odot$, $Y=s/R_\odot$, $Q=\sqrt {1/q^2 -1}$, and
$\xi=[ (\alpha^2+1)(1+Q^2 \sin^2 b) - \mu^2 ]^{1/2}$. Equation (\ref{angle}) recovers equation (9b) of
Adams \& Walker (1990) for a spherical halo ($q=1$).

The maximum extension $R_{max}$ of the halo is determined by the total mass $M_h$ and the asymptotic velocity $v_{rot}$ of the considered halo:

\begin{eqnarray}
R_{max}= G {M_h \over v_{rot}^2} ({ \sqrt {1-q^2} \over q\, {\rm Arccos}\, q})^{-1},
\label{rmax}
\end{eqnarray}

\noindent where $ v_{rot}\approx 220$ km s$^{-1}$ and $M_h\approx 2\times 10^{12}\,\msol$ for the Milky Way.

A first estimate of the diffuse emission can be obtained by assuming a Dirac function for the IMF:
$ dN/d\mwd=N_0\times \delta(m-m_0)$, where $N_0$ is fixed by the normalization condition $N_0=X_{WD}\times (M_h/m_0)$.
In that case the specific luminosity simply reads :

\begin{eqnarray}
\Gamma_\nu= { L_\nu (m_0,\tau) \over m_0 }
\end{eqnarray}

\begin{table}
\caption[]{Magnitude (in mag/arcmin$^2$), top row, and surface brightness $I_\nu$ (in mJy/str), bottom row, for a remnant WD population with $X_{wd}=1\%$ in the Milky Way halo at $(b,l)=(45^0,0^0$),
for a Dirac IMF. The flux and magnitudes are centered at $\lambda_0$ and averaged over $\Delta \lambda$ (in $\mu$m).}
\bigskip
\begin{tabular}{lcccccccc}
\tableline
Filter & B & V & R & I & J & H & K & M \\
$\lambda_0$ $[\mu$m$]$ & 0.44 & 0.55  & 0.70 & 0.90 & 1.25  & 1.65 & 2.20 & 5.0 \\
$\Delta \lambda$  $[\mu$m$]$ & 0.10 & 0.09  & 0.22 & 0.24 & 0.30  & 0.35 & 0.40 & 0.3 \\
\hline \\
$m_{wd}=0.6\,\msol$  & 28.7 & 27.6 & 27 & 26.4 & 25.9 & 26 & 26 & 25.6 \\
                                        & 180 & 374 & 531 & 742 & 890 & 307 & 280 & 121\\
$m_{wd}=0.8\,\msol$  & 29.6 & 28.5 & 27.8 & 27.2 & 27 & 27.1 & 27.2 & 26.7 \\
                                        & 81 & 171 & 264 & 351 & 343 & 107 & 93 & 44 \\
\tableline
\end{tabular}
\end{table}

\begin{figure}
\plotone{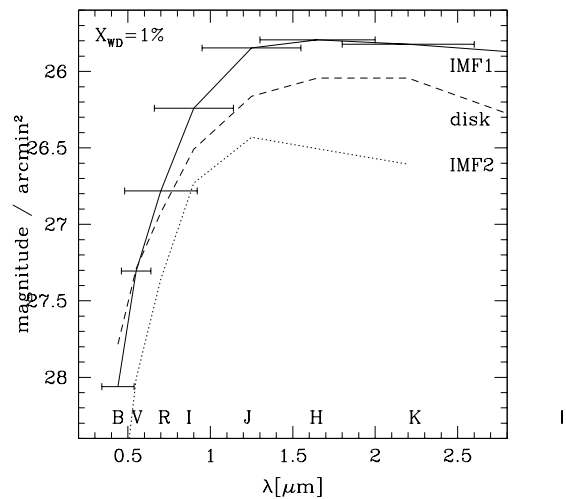}
\caption{Surface brightness of the Milky Way halo containing a WD fraction $X_{WD}=1\%$ of the dark matter density in different passband filters for three different IMFs, as described in the text.
\label{f3.ps}}
\end{figure}

Table I displays the expected surface brightness and apparent magnitudes of the WD diffuse emission
in the Milky Way dark halo, at $(b,l)=(45^0,0^0)$, for such a Dirac IMF with $m_0=0.6\,\msol$ and $m_0=0.8\,\msol$, respectively, for $X_{WD}=1\%$ and $t_h = 13$ Gyr.
Figure \ref{f3.ps} illustrates the apparent magnitudes of the WD diffuse emission for the same conditions for different IMFs, in various optical and near-infrared bands. The mass functions IMF1 and IMF2 refer to the WD-dominated IMFs of Chabrier et al. (1996) (eqn.(\ref{imf})) mentioned in \S 2. The IMF PopIII+II.5
displayed in Figure 1 gives results similar to IMF2. The IMF labeled "disk" refers to the IMF of the present Galactic disk (Chabrier 2003, Table I). The dependence of the flux upon the IMF between IMF1, which yields $\langle m_{WD}\rangle \simeq 0.6\,\msol$, and IMF2,
which yields $\langle m_{WD}\rangle \simeq 0.7\,\msol$, illustrates the strong dependence of WD cooling upon the WD mass (see Chabrier et al. 2000), whereas the difference between IMF1 and IMF "disk", which both have similar WD average mass, stems from the larger average WD luminosity $\langle L_\nu \rangle$ in the case of IMF1.
Note from equation (\ref{angle}) that the received radiation
has an angular dependence: magnitudes are found to increase by about 1.5 mag from $b=0^0$ to $b\ge 90^0$ and by about 1 mag from $l=0^0$ to $l= 180^0$. As mentioned above, the magnitudes have been calculated in standard filters, normalized to Vega. The flux of Vega decreases monotonically from $\sim 0.5\,\mu$m (V-band) redward, yielding as a spurious effect decreasing WD magnitudes redward of the J band. The flux, however, peaks in the I and J bands, as seen in Table I. These bands thus seem to be favored for the observation of the remnant WD surface brightness. Since $I_\nu$ is proportional to $X_{WD}$ (eqn.(\ref{bright})),
these curves can be easily scaled to different values of the expected halo WD mass fraction.
Although difficult to detect with present
instruments, the detection of such a halo WD emission might be observable with
the HST or with the JWST future satellite mission. We will come back to this point later. An other possibility, as explored in the next section, is the observation of halos of external galaxies.

\subsection {External galaxy halo}

As seen from eqn.(\ref{bright}), the surface brightness does not depend on the distance, because of the canceling $r^2$ dependence of the flux density and volume integration.
We thus calculate now the expected surface brightness of our primordial remnant population for the halo of an external galaxy, located at distance $d$ from Earth. The galactic density distribution is supposed to obey the same relation as given by eqn.(\ref{rho}):

\begin{eqnarray}
\rho(R,z)= \rho^\prime_{h_\odot} { 1 \over R^2 + a^{\prime^2} + z^2/q^{\prime^2}}
\label{rho2}
\end{eqnarray}

 We assume for this galaxy a rotation curve, and thus a dynamical halo identical to the Milky Way conditions, with a similar normalization $\rho^\prime_{h_\odot}=\rho_{h_\odot}\times (R_\odot^2 + a^{\prime^2})$, $a^\prime=a$, $q^\prime=q$. 
Such conditions apply for example to the spiral galaxy NGC5907 (Lequeux et al. 1998, Figure 2). 

\begin{figure}
\plotone{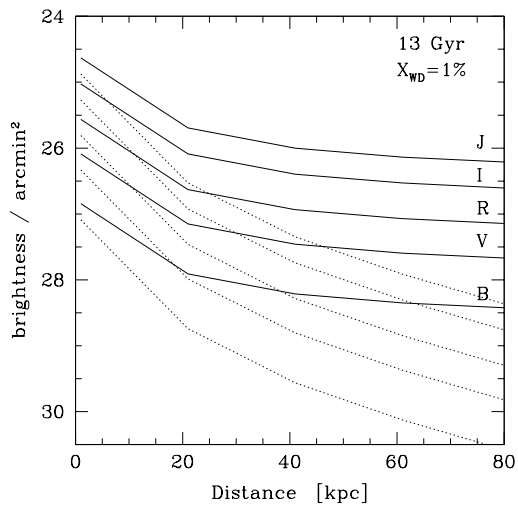}
\caption{Surface brightness radial profile from galactic center in different bands for an external galactic halo composed of a mass fraction $X_{WD}=1\%$ of primordial WDs, including (solid line) and without (dotted line) the Milky Way halo contribution.
\label{f4.ps}}
\end{figure}

We note $x,y,z$ the cartesien galactocentric coordinates of the external galaxy, with $z$ the direction perpendicular to the plane of the galactic disk. 
Galaxies seen edge-on are more favorable targets since in that case observations far away from the galactic center are not polluted by the light from the disk. In that case,
the surface brightness $I^\prime_\nu$ observed at  Earth now reads, with $x$ denoting the distance along the Earth-galaxy line of sight and $V_{obs}$ the observed galactic volume:

\begin{eqnarray}
I^\prime_\nu={ f^\prime_\nu \over \Omega }&=&{1\over \Omega} \int_{V_{obs}} dV\,\rho(x,y,z,)\,{\Gamma_\nu \over 4\pi s^2}\nonumber \\
&=&\int_{-x^\prime}^{+x^\prime}dx\,(d+x)^2\,\rho(x,y,z,)\,{\Gamma_\nu \over 4\pi (d+x)^2}\nonumber \\
&=& {\Gamma_\nu \over 4\pi}\rho_{h_\odot}^\prime {\mathcal I}^\prime(y,z),
\label{bright2}
\end{eqnarray}

\noindent with

\begin{eqnarray}
{\mathcal I}^\prime(y,z)&=&\int_{-\sqrt{R^{{\prime}^2}_{max}-y^2}}^{+\sqrt{R^{{\prime}^2}_{max}-y^2}} dx\,{1\over x^2+y^2+z^2/q^{\prime^2}+a^{\prime^2}}\nonumber \\
 &=& {2\over \sqrt {y^2+z^2/q^{\prime^2}+a^{\prime^2}}}
\,\,tg^{-1} \Bigl[ {\sqrt{R^{{\prime}^2}_{max}-y^2} \over \sqrt {y^2+z^2/q^{\prime^2}+a^{\prime^2}}}\Bigl],\nonumber \\
\end{eqnarray}

\noindent and the maximum halo extension $R^\prime_{max}$ is fixed by the halo mass, as in eqn.(\ref{rmax}). For $M_h\sim 2\times10^{12}\,\msol$, $R^\prime_{max}\sim 150$ kpc, so that for galaxies situated at
distances $d\sim$10-100 Mpc, the halos extend to angular sizes $\sim 0.1^{0}$-1$^{0}$.
Figure \ref{f4.ps}  portrays the expected radial dependence in the galactic plane of the halo of the galaxy NGC5907 alone ($(b,l)=(51.1^0,91.6^0)$) and of the cumulative contributions of this halo plus our own halo, for $X_{wd}=1\%$, in various passbands. The total flux values are given in Table II. Depending on the radial distance $y$, our own halo contributes from $\sim 20\%$ near the galactic center to $\sim$ 80\% at 80 kpc to the total flux. The diffuse emission is below the detection limit
of 
present large, deep-field cameras like MEGACAM or WIRCAM,
but should be in reach of space-based HST or JWST deep field surveys.
Note that both the
angular modulation and radial dependence of the signal will help separate the halo radiation from other possible sources, helping correcting for the sky level contribution.
Present observations of NGC5907 (Lequeux et al. 1996, Rudy et al. 1997, Zheng et al. 1999) extend only to 100$^{\prime\prime}$ from the galactic center, i.e. $\sim 7$ kpc for a distance $d=14$ Mpc. At this distance, contamination from the disk and the bulge is not negligible.

\begin{table}
\caption[]{Same as Table 1 for the cumulative contributions of the Milky Way halo and a similar external halo (namely NGC5907) at  40 kpc from the galactic center, for the IMF1 and $X_{WD}=1\%$. At this distance, the Milky Way halo contributes $\sim$ 70\% to the total flux.}
\bigskip
\begin{tabular}{lcccccccc}
\tableline
& B & V & R & I & J & H & K & M  \\
\hline \\
mag  & 28.2 & 27.4 & 26.9 & 26.4 & 26.0 & 25.9 & 26.0 & 26.2 \\
$I_\nu$ (mJy/str)   &  284 & 454 & 570 & 732 & 831 & 314 & 302 & 71 \\
\tableline
\end{tabular}
\end{table}

\begin{figure}
\plotone{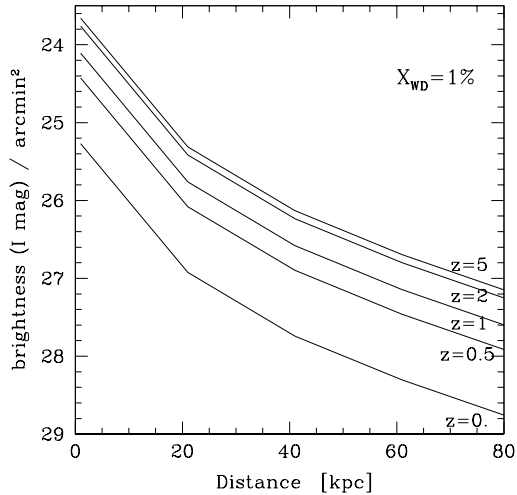}
\caption{ I-band surface brightness radial profile for an external galactic halo with a fraction $X_{WD}=1\%$ of primordial WDs for different redshift values.
\label{f5.ps}} 
\end{figure}

\begin{figure}
\plotone{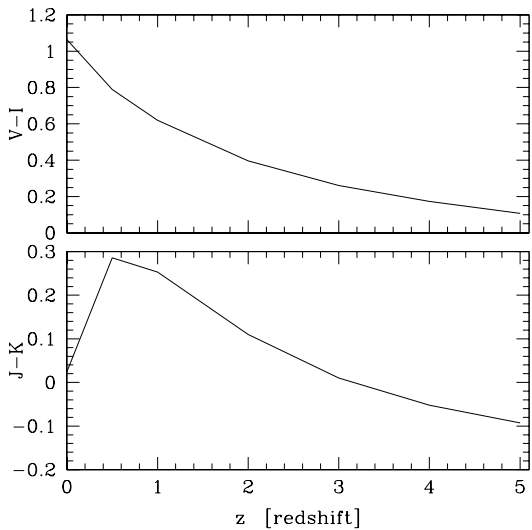}
\caption{Color dependence as a function of redshift for a galactic halo composed of a mass fraction $X_{WD}=1\%$ of primordial WDs.
\label{f6.ps}} 
\end{figure}

Alternatively, an indication on the primordial WD population of dark halos would be their contribution at high redshift. High redshift means younger ages and thus brighter WD luminosities. Figure \ref{f5.ps}  displays the redshift dependence of the halo WD I-band surface brightness, for various redshift values up to $z=5$, i.e. $t_h \sim 1$ Gyr, for $X_{WD}=1\%$. Already at $z=1$, $t_h\sim 6$ Gyr, the remnant emission is about 1 magnitude brighter than at $z=0$. At high redshift, however, some of the WD progenitors are still on the main sequence or the AGB phase and contribute significantly to the background light. A fully consistent calculation requires chemico-evolution codes (see e.g. Lee et al. 2004). Appropriate color-cuts, however, might be able to observe this contribution from halo WDs at larger redshifts. Figure \ref{f6.ps} illustrates the color dependence of the WD halo signature under the same conditions. 

\section{Conclusion}

In this paper, we have examined the possibility for a fraction of the baryons in the universe to be trapped in the WD remnant population of a primordial generation of intermediate or low-mass stars, formed before the end of the reionization period, i.e. in the redshift range $5\simle z \simle 20$. For such redshift values, a non-negligible fraction of the nucleosynthetic
baryons can be in the form of such WD progenitors without producing unacceptable light today. As a possible signature of such relics, we calculate their expected surface brightness both in our own halo and in external galactic halos. The expected radiation lies below the detection limits of
ground-based present surveys,
but the detection of this glow should
be in reach of the HST or the JWST, in particular if several
external galactic halos can be observed along a given line-of-sight. Both a successful or a null detection of such a diffuse light, within relevant detection limits, will provide precious information about the mode of star formation and the
evolution of the IMF at early stages of galaxy formation. The background light contribution of these WDs is found to increase appreciably with redshift, suggesting possible detection in dedicated surveys, with appropriate color cuts. Contributions from red giants and main sequence stars at these redshifts, however, must be calculated consistently with chemico-evolution models, in order to examine the favored detection bands. 

\acknowledgments The author is indebted to the Erskine foundation and the department of physics and astronomy of the university of Canterbury, where this work was completed, for their wonderful hospitality. Thanks also to J. Devriendt, P. Ferruit, J. Lequeux, T. Forveille and M. Rieke for helpful conversations
and to the anonymous referee for highly valuable comments.


\begin{references}

Abel, T., Bryan, G., \& Norman, M., 2000, \apj, 540, 39

Abel, T., Anninos, P., Norman, M., \& Zhang, Y., 1998, \apj, 508, 518

Adams, F., \& Walker, T., 1990, \apj, 359, 57

Afonso, C., et al., 2003, \aap, 400, 951

Alcock, C., et al., 2000, \apj, 542, 281

Allen, C.W., 1991, Astrophysical quantities, Athlone Press

Barkana, R., \& Loeb, A., 2001, Phys. Rep., 349, 125

Bessell, M., \& Brett, C., 1988, \pasp, 100, 1134

Bromm, V., \& Larson, R., 2004, \araa, to appear

Bromm, V., Coppi, P., \& Larson, R., 1999, \apj, 527, L5

Bromm, V., Coppi, P., \& Larson, R., 2002, \apj, 564, 23

Brook, C., Kawata, D., \& Gibson, B., 2003, \mnras, 343, 913

Brook, C., \& Loeb, A., 2003, Nature, 425, 812

Chabrier, G., Segretain, L., \& M\'era, D., 1996, \apjl, 468, L21

Chabrier, G., 1999a, \apjl, 513, L103

Chabrier, G., 1999b, in {\it The Third Stromlo Symposium: The Galactic Halo}, ASP Conf. Ser. 165, 399

Chabrier, G., Brassard, P., Fontaine, G., \& Saumon, D., 2000, \apj, 543, 216

Chabrier, G., 2003, \pasp, 115, 763

Christlieb, N., et al., 2002, Nature, 419, 904

Clarke, C., \& Bromm, V., 2003, \mnras, 343, 1224

Fields, B., Freese, K., \& Graff, D., 2000, \apj, 534, 265

Flynn, C., Holopainen, J., \& Holmberg, J., 2003, \mnras, 339, 817

Gould, A., Flynn, C., \& Bahcall, J.N., 1998, \apj, 503, 798

Hansen, B, 2001, \apj, 558, L39

Hauser, G., \& Dwek, E, 2001, \araa, 39, 249

Larson, R., 1986, \mnras, 218, 409

Lee, H-C, Gibson, B., Fenner, Y., Kawata, D., \& Renda, A., 2004, in {\it The 5h Workshop on Galactic Chemodynamics}, Publ. Astr. Soc. Australia, to be published

Lequeux, J., Fort, B., Dantel-Fort, M., Cuillandre, J.-C., \& Mellier, Y., 1996, \aap, 312, L1

Lequeux, J., Combes, F., Dantel-Fort, M., Cuillandre, J.-C., Fort, B., \& Mellier, Y., 1998, \aap, 334, L9

Mackey, J., Bromm, V., \& Hernquist, L., 2003, \apj, 586, 1

Madau, P., \& Pozzetti, L., 2000, \mnras, 312, L9

Marigo, P., Chiosi, C., Girardi, L. \& Wood, P., 2001, \aap, 371, 152

Magliocchetti, M., Salvaterra, R., \& Ferrara, A., 2003, \mnras, 342, L25

Nakamura, F., \& Umemura, M., 2001, \apj, 548, 19

Nakamura, F., \& Umemura, M., 2002, \apj, 569, 549

Padoan, P., \& Nordlund, \aa A, 2002, \apj, 576, 870

Omukai, K., \& Yoshii, Y., 2003, \apj, 599, 746

Oppenheimer B., et al., 2001, {\it Science}, 292, 698

Portinari, L., Moretti, A., Chiosi, C., \& Sommer-Larsen, J., 2004, \apj, 604, 579

Reid, I.N., Sahu, K., \& Hawley, S., 2001, \apj, 559, 942

Reyl\'e, C., Robin, A., \& Cr\'ez\'e, M., 2001, \aap, 378, L53

Rudy, R., Woodward, C., Hodge, T., Fairfield, S., \& Harker, D., 1997, Nature, 387, 159

Salvaterra, R., \& Ferrara, A., 2003, \mnras, 339, 973

Salvaterra, R., Ferrara, A., \& Schneider, R., \mnras, submitted, astro-ph/0304074

Santos, M., Bromm, V., \& Kamionkowski, M., 2002, \mnras, 336, 1082

Saumon, D., \& Jacobson, 1999, \apj, 511, L107

Schaerrer, D., 2002, \aap, 382, 28

Stiavelli, M., Fall, M., \& Panagia, N.,  2004, \apj, 600, 508

Venkatesan, A., \& Truran, J., 2003, \apj, 594, L1

White, S.D.M. \& Springel, V., In {\it The first stars}, Eds. A. Weiss, T. Abel \& V. Hill, Springer, p. 327

Zheng, Z., et al., 1999, \aj, 117, 2757

\end{references}
\end{document}